\documentclass[aps,prl,twocolumn,superscriptaddress,groupedaddress,amsmath,amssymb]{revtex4-1}
\usepackage{graphicx}
\usepackage{dcolumn}
\usepackage{bm}
\usepackage{xcolor}
\usepackage{float}
\usepackage{booktabs}
\usepackage{soul}
\definecolor{aa}{RGB}{0,0,139}

\newcommand{\BESIIIorcid}[1]{\href{https://orcid.org/#1}{\hspace*{0.1em}\raisebox{-0.45ex}{\includegraphics[width=1em]{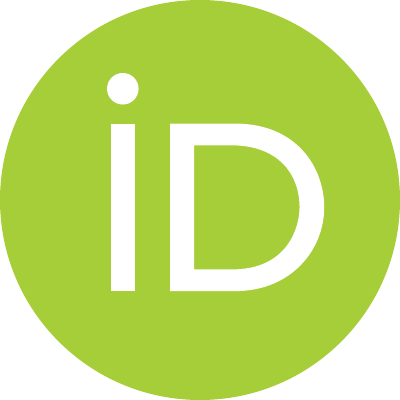}}}}

\usepackage[colorlinks,urlcolor=aa,linkcolor=blue,anchorcolor=aa,citecolor=aa]{hyperref}

\usepackage[mathlines]{lineno}
\usepackage{multirow}
\usepackage{enumerate}
\usepackage{relsize}
\usepackage{makecell}
\def\babar{\mbox{\slshape B\kern-0.1em{\smaller A}\kern-0.1em
    B\kern-0.1em{\smaller A\kern-0.2em R}}}
\makeatletter
\renewcommand\footnote[1]{%
    \stepcounter{footnote}%
    \protected@xdef\@thefnmark{\thefootnote}%
    \@footnotemark
    \@footnotetext{#1}%
}
\makeatother

\begin{document}


\preprint{APS/123-QED} 
\title{{\bf \boldmath Observation of $D^+ \to K^{*0}\rho^+$ and $D^+\to K^{*+}\rho^0$ in Doubly Cabibbo-Suppressed Decay $D^+ \to K^+\pi^+\pi^-\pi^0$}}

\author{
\begin{small}
\begin{center}
M.~Ablikim$^{1}$\BESIIIorcid{0000-0002-3935-619X},
M.~N.~Achasov$^{4,d}$\BESIIIorcid{0000-0002-9400-8622},
P.~Adlarson$^{81}$\BESIIIorcid{0000-0001-6280-3851},
X.~C.~Ai$^{87}$\BESIIIorcid{0000-0003-3856-2415},
C.~S.~Akondi$^{31A,31B}$\BESIIIorcid{0000-0001-6303-5217},
R.~Aliberti$^{39}$\BESIIIorcid{0000-0003-3500-4012},
A.~Amoroso$^{80A,80C}$\BESIIIorcid{0000-0002-3095-8610},
Q.~An$^{77,64,\dagger}$,
Y.~H.~An$^{87}$\BESIIIorcid{0009-0008-3419-0849},
Y.~Bai$^{62}$\BESIIIorcid{0000-0001-6593-5665}
% ...
\\
\vspace{0.2cm}
(BESIII Collaboration)\\
\vspace{0.2cm} {\it
$^{1}$ Institute of High Energy Physics, Beijing 100049, People's Republic of China\\
% ...
$^{4}$ Budker Institute of Nuclear Physics SB RAS (BINP), Novosibirsk 630090, Russia\\
% ...
$^{31}$ INFN Sezione di Ferrara, (A)INFN Sezione di Ferrara, I-44122, Ferrara, Italy; (B)University of Ferrara, I-44122, Ferrara, Italy\\
% ...
$^{39}$ Johannes Gutenberg University of Mainz, Johann-Joachim-Becher-Weg 45, D-55099 Mainz, Germany\\
% ...
$^{62}$ Southeast University, Nanjing 211100, People's Republic of China\\
% ...
$^{64}$ State Key Laboratory of Particle Detection and Electronics, Beijing 100049, Hefei 230026, People's Republic of China\\
% ...
$^{77}$ University of Science and Technology of China, Hefei 230026, People's Republic of China\\
% ...
$^{80}$ University of Turin and INFN, (A)University of Turin, I-10125, Turin, Italy; (B)University of Eastern Piedmont, I-15121, Alessandria, Italy; (C)INFN, I-10125, Turin, Italy\\
$^{81}$ Uppsala University, Box 516, SE-75120 Uppsala, Sweden\\
% ...
$^{87}$ Zhengzhou University, Zhengzhou 450001, People's Republic of China\\
\vspace{0.2cm}
$^{\dagger}$ Deceased\\
% ...
$^{d}$ Also at the Novosibirsk State University, Novosibirsk, 630090, Russia\\
% ...
}
\end{center}
\vspace{0.4cm}
\end{small}
}

\vspace{0.4cm}
\begin{abstract}
By analyzing an $e^+e^-$ collision data sample with an integrated luminosity of 20.3 fb$^{-1}$ collected with the BESIII detector at the center-of-mass energy of 3.773 GeV, we perform the first amplitude analysis on the doubly Cabibbo-suppressed decay $D^+ \to K^+\pi^+\pi^-\pi^0$ and report the first observation of $D^+ \to K^{*0}\rho^+$ and $D^+\to K^{*+}\rho^0$.  The corresponding branching fractions are $(5.67\pm0.41_{\rm stat}\pm0.17_{\rm syst})\times10^{-4}$ and $(5.32\pm0.57_{\rm stat}\pm0.24_{\rm syst})\times10^{-4}$, respectively. These two $D\to VV$ decay both have large transverse polarizations. The longitudinal polarization fractions are measured to be $0.111\pm0.024_{\rm stat}\pm0.008_{\rm syst}$ and $0.263\pm0.049_{\rm stat}\pm0.015_{\rm syst}$, respectively. The branching fraction of the decay $D^+\to K^+\omega$ is measured to be $(4.76\pm0.84_{\rm stat}\pm0.13_{\rm syst})\times 10^{-5}$.
\end{abstract}

\pacs{Valid PACS appear here}
\maketitle

The charm quark in a hadronic charmed-hadron decay predominantly undergoes a Cabibbo-favored (CF) transition, $c \to s W^+$, with the $W^+$ converting to a $u\bar{d}$ pair. The corresponding decay amplitudes scale as $V_{cs}^*V_{ud}$, where $V_{ij}$ denotes the element of CKM matrix. In doubly Cabibbo-suppressed (DCS) decays, both $c$ and $W^+$ transitions are suppressed, $c \to d$ and $W^+ \to u\bar{s}$, leading to amplitudes proportional to $V_{cd}^*V_{us}$. This gives an expectation that the branching fraction (BF) ratio for DCS and CF modes sharing identical weak topologies is $|V_{cd}^*V_{us}/V_{cs}^*V_{ud}|^2 = \tan^4\theta_C \approx 0.29\%$~\cite{Lipkin:2002za,Cheng:2010ry}, where $\theta_C$ is the Cabibbo angle.
However, deviations from this expectation can arise from SU(3) symmetry breaking effects or long-distance weak interaction topologies. 
On the other hand, the validity of SU(3) symmetry is well established for CF decays, while significant SU(3) symmetry breaking has been observed in singly Cabibbo-suppressed decays such as $D^0\to\pi^+\pi^-/K^+K^-$. Indeed, this symmetry breaking may help to explain the observed $C\!P$ violation in such decays.  
DCS decays could exhibit an even larger contribution from SU(3) symmetry breaking due to their smaller amplitude scale, yet they remain largely unexplored. Therefore, detailed studies of DCS decays, including examining decay topologies and polarization dynamics, are crucial to test whether their underlying mechanisms are consistent with those of CF processes, thereby probing SU(3) symmetry breaking and non-perturbative QCD.

\begin{figure}[tp!]
\begin{center}
  \begin{minipage}[t]{8.8cm}
      {\includegraphics[width=1.\linewidth]{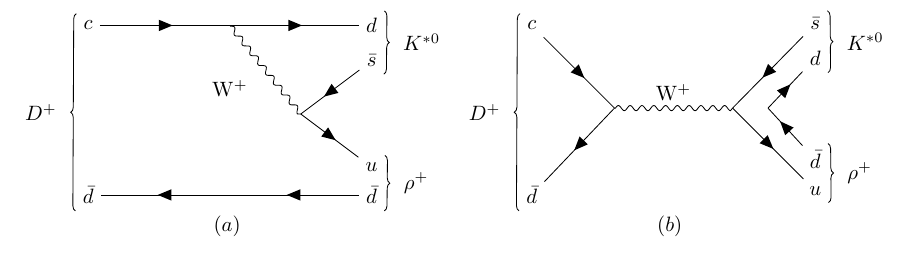}}
      {\includegraphics[width=1.\linewidth]{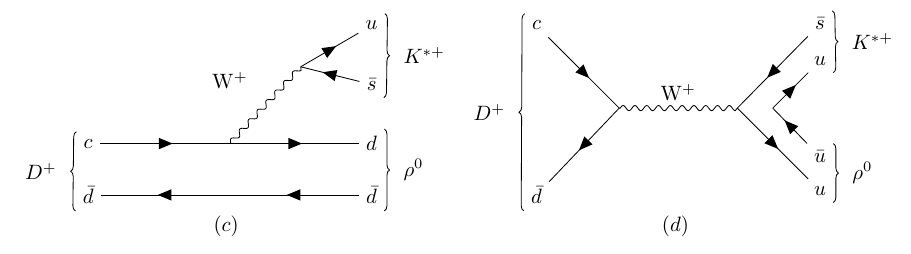}}
      \\
   \end{minipage}

   \caption{Topological diagrams of (a,b) $D^+\to K^{*0} \rho^+$ and (c,d) $D^+\to K^{*+}\rho^0$ decays. The topologies are (a) internal $W$ emission, (c) external $W$ emission, and (b,d) $W$ annihilation.}
\label{topologya0}
\end{center}
\end{figure}

The description of the decay dynamics of charmed mesons relies on a model-independent diagrammatic approach to decompose the decay amplitude into various topological amplitudes, i.e.,~external $W$ emission~($T$), internal $W$ emission~($C$), $W$ exchange in $D^0$ decays~($E$) and $W$ annihilation in $D^+$ decays~($A$).  Note that penguins ($P$) do not contribute to strangeness-changing charm decays.  
In general, the hierarchical pattern $|T|>|C|>|E|>|A|$ is well established in CF $D\to PP$ and $D\to VP$ decays~\cite{Cheng:2024hdo}, but some exceptions have been observed in DCS decays. Here, $P$ and $V$ denote pseudo-scalar and vector mesons, respectively. For instance, $D^+\to K^+\omega$ proceeds through $T$ and $A$ amplitudes. Its significantly suppressed BF~(i.e., $(5.7^{+2.5}_{-2.1~\rm stat}\pm0.2_{\rm syst})\times 10^{-5}$)~\cite{BESIII:2020wnc} 
might suggest $T$ and $A$ have comparable magnitudes, or that there is a significant contribution from the flavor-singlet topological amplitude.
Similarly, while exact SU(3) symmetry predicts identical topological amplitudes between CF and DCS decays, studies of $D^0\to VP$ decays have revealed that the CF and DCS channels exhibit significant differences in both $T$ and $E$ amplitudes, implying very large SU(3) breaking effects~\cite{Cheng:2016ejf,Cheng:2024hdo}. Whether such breaking persists in the $D\to VV$ sector remains an open question.
The topological amplitudes of the DCS decays $D^+\to K^{*0}\rho^+$ and $D^+\to K^{*+}\rho^0$, accessible via amplitude analysis of $D^+\to K^+\pi^+\pi^-\pi^0$, offer a unique probe for testing this breaking.
Their distinct topological amplitudes, ($C+A$) and ~($T-A$) shown in Fig.~\ref{topologya0}, enable the isolation of the weak-annihilation amplitude $A$, allowing us to directly examine SU(3) breaking in the $D\to VV$ sector.

Polarization properties of $VV$ final states provide another powerful tool to understand the underline mechanism of CF and DCS decays.
Some polarization puzzles have been raised in measurements of CF decays, but are rarely discussed in DCS decays~\cite{Cao:2023csx}. Current CF measurements reveal that $D^0\to K^{*-}\rho^+, \bar{K}^{*0}\rho^0$ are dominated by $\mathcal{D}$-wave~\cite{BESIII:2017jyh,BESIII:2019lwn}, whereas $D^+ \to \bar{K}^{*0}\rho^+$ is primarily $\mathcal{S}$-wave~\cite{BESIII:2023qgj,BESIII:2024waz}. The key dynamical difference of the above decays is that the $D^0$ decays receive additional contribution from $E$, which appears to invert the naive expectation of polarization hierarchy $|\mathcal{S}|>|\mathcal{P}|>|\mathcal{D}|$~\cite{Cheng:2024hdo}. Other measurements also find the $D^0 \to \rho^+\rho^-, \rho^0\rho^0$~\cite{BESIII:2023exz} and $D_s^+ \to \omega\rho^+$~\cite{BESIII:2025owp} decays are dominated by $\mathcal{D}$-wave, and $D_s^+ \to K^{*0}\rho^+, K^{*+}\rho^0$ decays by $\mathcal{P}$-wave~\cite{BESIII:2022bvv}. The measurement of DCS decays $D \to K^*\rho$ will test polarization predictions and quantify the weak-annihilation contribution in the DCS sector.

In addition, the $D^+\to K^+\pi^+\pi^-\pi^0$ decay include several intermediate $D\to AP$ amplitudes, which will provide unique information to study axial-vector mesons in DCS decays, e.g., the $a_1(1260)$ and $K_1(1270,1400)$~\cite{Cheng:2011pb,Cheng:2013cwa,Guo:2018orw}.

\nolinenumbers

This Letter reports the first amplitude analysis of the DCS decay $D^+ \to K^+\pi^+\pi^-\pi^0$ using 20.3~fb$^{-1}$ of $e^+e^-$ collision data~\cite{BESIII:2024lbn} collected with BESIII detector at the center-of-mass energy of $\sqrt{s}$=3.773~GeV. The longitudinal polarization fractions of $D^+\to K^{*0}\rho^+$ and $D^+\to K^{*+}\rho^0$ are also measured for the first time. Benefiting from the resonance peak of $\psi(3770)$ near this energy, the $D\bar{D}$ pairs are produced abundantly without accompanying hadron(s), thereby providing an ideal environment for studying $D$ meson decays with the double-tag (DT) technique, pioneered by the Mark III Collaboration~\cite{MARK-III:1985hbd}. Charge-conjugate decays are always implied throughout this letter.

A detailed description of the design and performance of the BESIII detector is given in Ref.~\cite{BESIII:2009fln}. The inclusive Monte Carlo (MC) samples, described in Refs.~\cite{GEANT4:2002zbu,Jadach:1999vf,Jadach:2000ir,Lange:2001uf,Ping:2008zz}, are used to determine detection efficiencies and to estimate backgrounds. The signal MC samples of $D^+\to K^+\pi^+\pi^-\pi^0$ are generated based on the amplitude model obtained in this work.

In the data samples, the signal $D^+\to K^+\pi^+\pi^-\pi^0$ decays and the tagged $D^-$ decays are reconstructed simultaneously, where two tag modes $D^-\to K^+\pi^-\pi^-$ and $D^-\to K^+\pi^-\pi^-\pi^0$ are used. The selection criteria for the final-state particles, including $K^\pm$, $\pi^\pm$ and $\pi^0$ are the same as those in Ref.~\cite{BESIII:2024waz}. The tagged $D^-$ mesons are selected using two variables, the energy difference 
\begin{equation}
    \Delta E_{\rm tag}\equiv E_{D^-}-E_{\rm beam},
\end{equation}
and the beam-constrained (BC) mass
\begin{equation}
    M_{\rm BC}^{\rm tag}\equiv \sqrt{E_{\rm beam}^2/c^4-|\Vec{p}_{D^-}|^2/c^2},
\end{equation}
where $E_{\rm beam}$ is the beam energy, and $\Vec{p}_{D^-}$ and $E_{D^-}$ are the momentum and the energy of the $D^-$ candidate in the $e^+e^-$ rest frame. Similar variables are defined for the signal side of DT events.  For the tag side, $M_{\rm BC}^{\rm tag}$ is required to be within [1.863, 1.877]~GeV/$c^2$ and $\Delta E_{\rm tag}$ within [$-0.025, 0.024$]([$-0.057, 0.046$])~GeV for $D^-\to K^+\pi^-\pi^-$($\pi^0$). The best DT candidate is selected with the minimum sum $\Delta E^2_{\rm sig}+\Delta E^2_{\rm tag}$.

To suppress background from $D^+\to K^+K_S^0\pi^0, K_S^0\to \pi^+\pi^-$ and $D^+\to K^+K^-\pi^+, K^-\to\pi^-\pi^0$ decays, events are rejected if the invariant mass of the $\pi^+\pi^-$ and $\pi^-\pi^0$ are within the range $[0.4675, 0.5175]$~GeV/$c^2$ and $[0.475, 0.500]$~GeV/$c^2$, respectively.
The $D^+ \to K^+\eta$ decay produces a very narrow $\eta$ peak on $M_{\pi^+\pi^-\pi^0}$ spectrum, with its observed line shape dominated by the mass resolution. Its interference with other contributions is expected to be small and localized around the nominal $\eta$ mass. This decay is therefore excluded from the amplitude analysis by requiring $M_{\pi^+\pi^-\pi^0}>0.57~\mathrm{GeV}/c^2$.
To reduce non-$D^+D^-$ background, the opening angle between the $D^+$ and $D^-$ candidates is required to be larger than 160$^\circ$. To further increase the signal purity, the $\Delta E_{\rm sig}$ is required to be in the range $[-0.04, 0.02]$~GeV.

\begin{figure}[htp]
  \centering
    \includegraphics[width=0.23\textwidth]{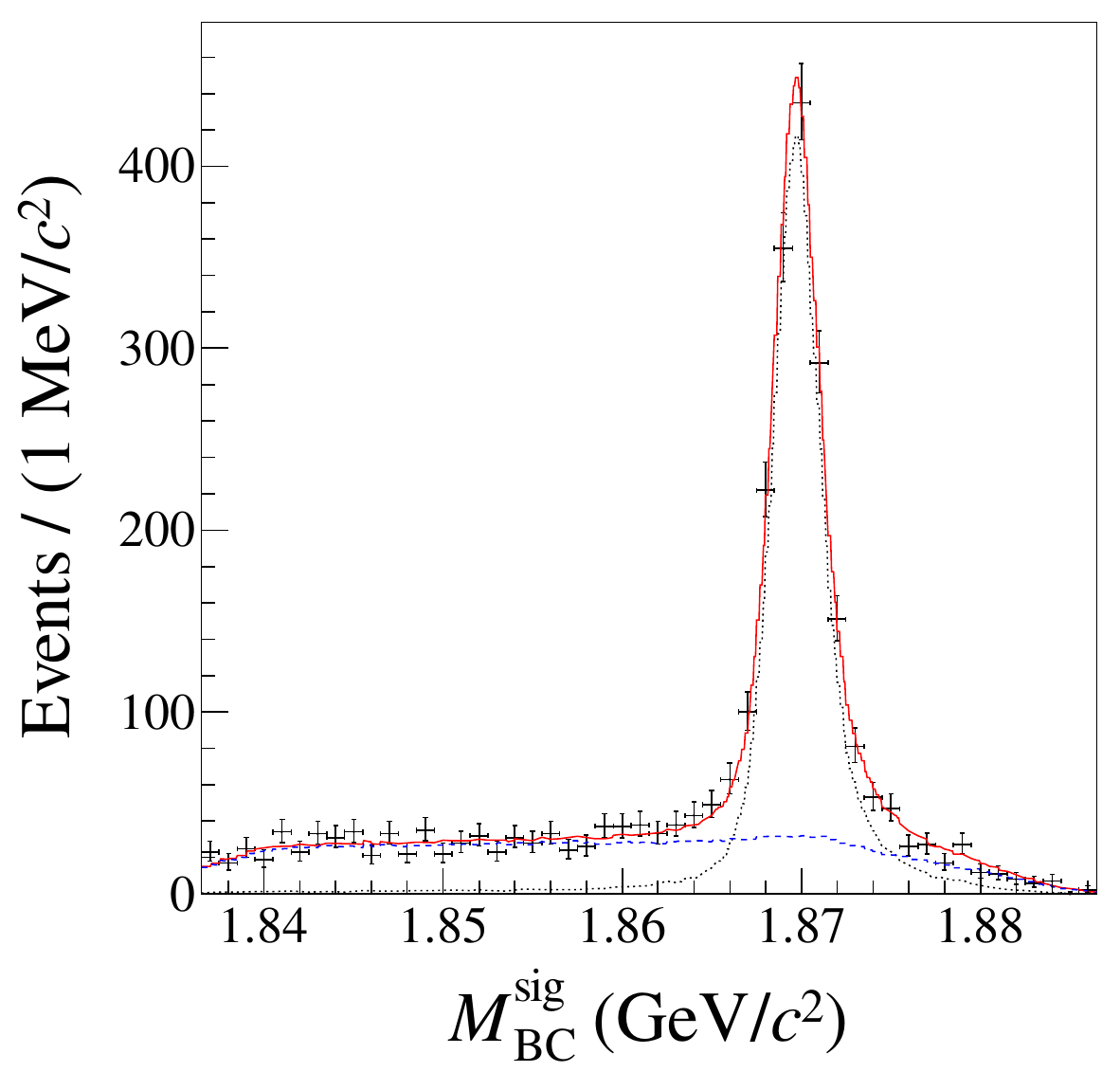}
    \includegraphics[width=0.23\textwidth]{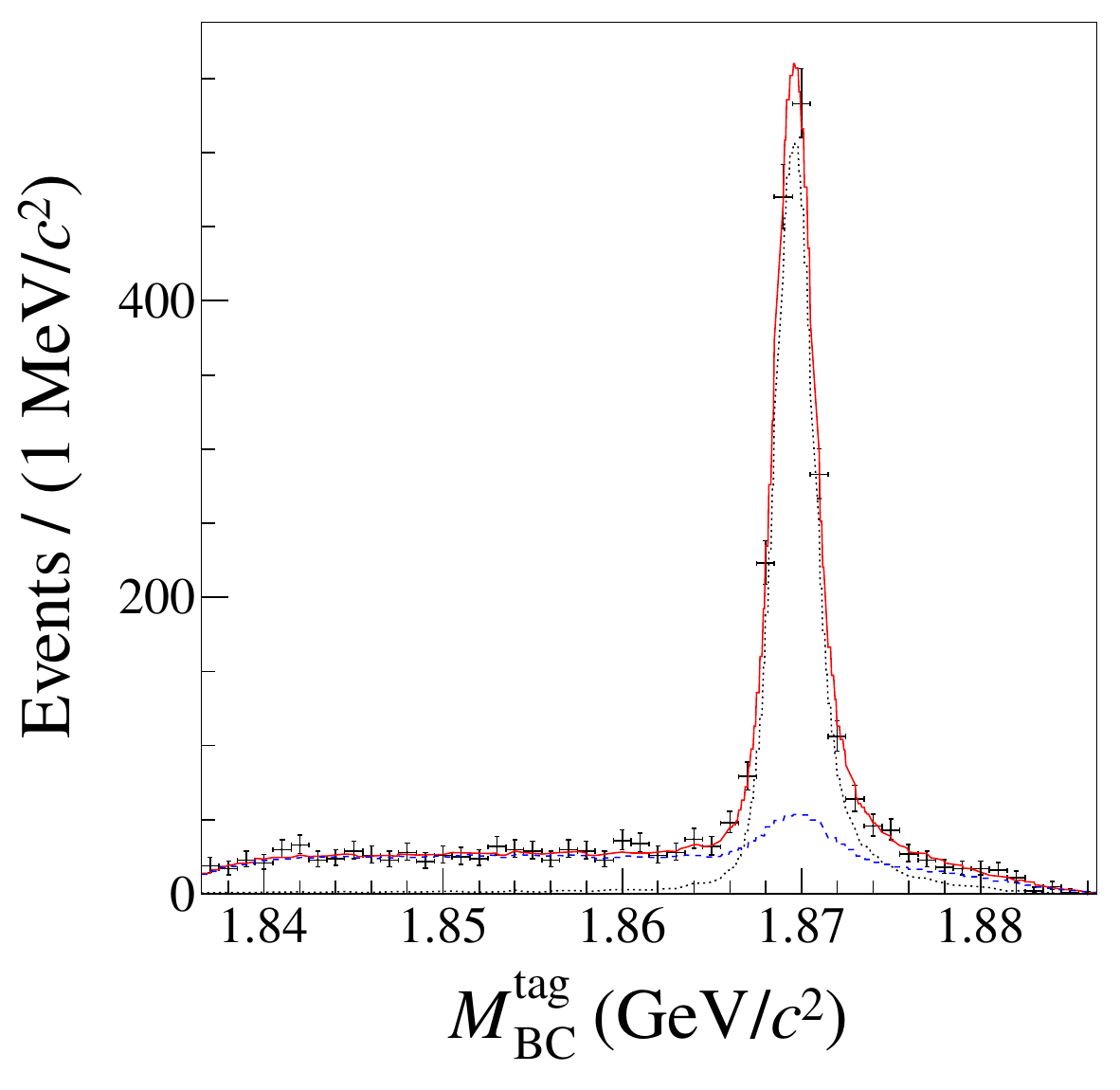}    
    \caption{Projections of the fit to the $M_{\rm BC}^{\rm sig}$ versus $M_{\rm BC}^{\rm tag}$ distribution of DT candidates. The data samples are represented by points with uncertainties, the total fit by the red line, the fitted signal by the black dotted line, the background by the blue dotted lines, respectively.}
    \label{BF:fitresult}
\end{figure}

In particular for the amplitude analysis, to ensure that all events are within the phase-space boundary and achieve a better resolution for the reconstructed momentum, an additional two-constraint kinematic fit is performed, constraining the masses of reconstructed signal $D^+$ and $\pi^0$ to the known values~\cite{pdg}.
With the requirements of $M_{\rm BC}^{\rm sig}\in[1.865, 1.875]$~GeV/$c^2$ and $M_{\rm BC}^{\rm tag}\in[1.863, 1.877]$~GeV/$c^2$, a total of 1739 DT events are selected for the amplitude analysis, with a signal purity of $\xi = (86.1\pm0.8)\%$. The purity and the DT yield are determined by a two-dimensional (2D) unbinned maximum-likelihood fit to the distribution of $M_{\rm BC}^{\rm sig}$ versus $M_{\rm BC}^{\rm tag}$, as shown in Fig.~\ref{BF:fitresult}. The signal and background shapes for the 2D probability density functions (PDF) are all extracted from the inclusive MC sample, except for a flat background which is described by a 2D ARGUS function~\cite{ARGUS:1990hfq}.
\nolinenumbers

An unbinned maximum likelihood method is performed for the amplitude analysis. The probability density function (PDF) depends on the four-momenta of the final-state particles ($\bm{x}$), with a combined signal and background model,
\begin{equation}
    \epsilon R_4 \, \left[ \xi\frac{|\mathcal{M}(\bm{x})|^2}{\int \epsilon|\mathcal{M}(\bm{x})|^2R_4d\bm{x}}+(1-\xi)\frac{B_{\epsilon}(\bm{x})}{\int \epsilon B_{\epsilon}(\bm{x})R_4d\bm{x}} \right],
\end{equation}
where $\epsilon$ denotes the acceptance function obtained from the signal MC events generated uniformly in the available phase space, $B_{\epsilon}(\bm{x})$ is defined as $B(\bm{x})/\epsilon$, and $R_4d(\bm{x})$ is the element of the four-body phase space. The background PDF $B(\bm{x})$ is estimated from the inclusive MC samples with the XGBoost technique~\cite{Rogozhnikov:2016bdp,Liu:2019huh}. The normalization integral is evaluated using the MC samples described in Ref.~\cite{Cranmer:2000du}.
The signal model is constructed as a coherent sum of intermediate processes $\mathcal{M}(\bm{x})=\sum\rho_ne^{i\phi_n}\mathcal{A}_n(\bm{x})$, where $\rho_ne^{i\phi_n}$ is the complex coefficient of the $n^{th}$ amplitude. The $n^{th}$ amplitude is defined as $\mathcal{A}_n(\bm{x}) = F_n^{1}F_n^{2}F_n^{D^+}S_nP_n^{1}P_n^{2}$, where the indices $i = 1, 2, D^+$ correspond to the two subsequent intermediate resonances and the $D^+$ meson, the $F_n^{i}$ represents the Blatt-Weisskopf barrier factors~\cite{Blatt_Weisskopf_1973}, the $P_n^i$ denotes the propagators, and $S_n$ is the spin factor describing the $\mathit{L-S}$ coupling in the amplitude, which is constructed using the covariant tensor formalism~\cite{Zou:2002ar}.

A relativistic Breit-Wigner (RBW) function~\cite{Jackson:1964zd} is employed to describe the propagators for the resonances $K^*$, $K_1(1270)$, $K_1(1400)$, $a_1(1260)$, $f_2(1270)$, and $\omega$. The $\rho$ resonance is characterized by the Gounaris-Sakurai line shape~\cite{Gounaris:1968mw}. The $f_0(500)$ resonance is described by a two-channel Flatte-like formula~\cite{Bugg:1996ki}. For $K_1(1270)$, $K_1(1400)$ and $a_1(1260)$, which are treated as quasi-three-body decays, the corresponding widths are determined by integrating the relevant squared amplitude over the phase space~\cite{dArgent:2017gzv}. The remaining masses and widths of the intermediate resonances used in the fit are taken from Ref.~\cite{pdg}. 

\begin{figure}[htp]
  \begin{center}
  \includegraphics[width=4.2cm]{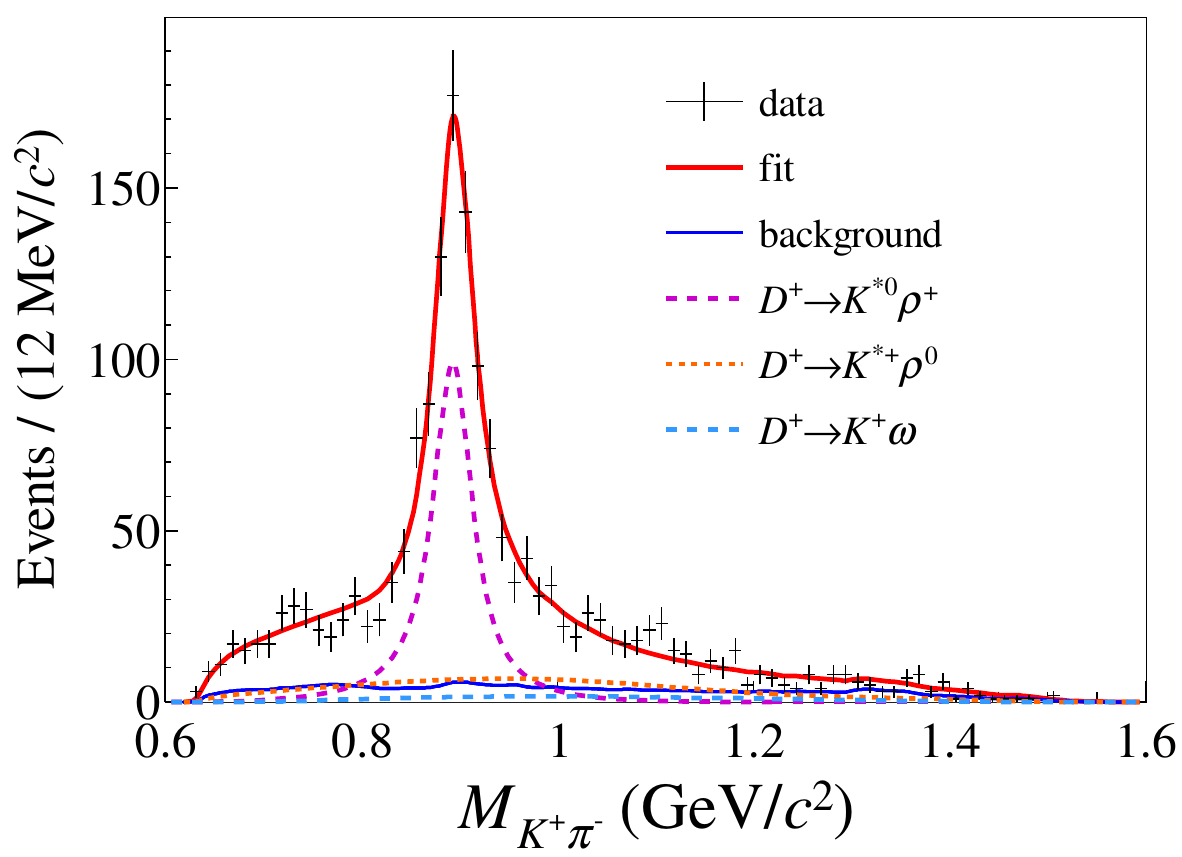}
  \includegraphics[width=4.2cm]{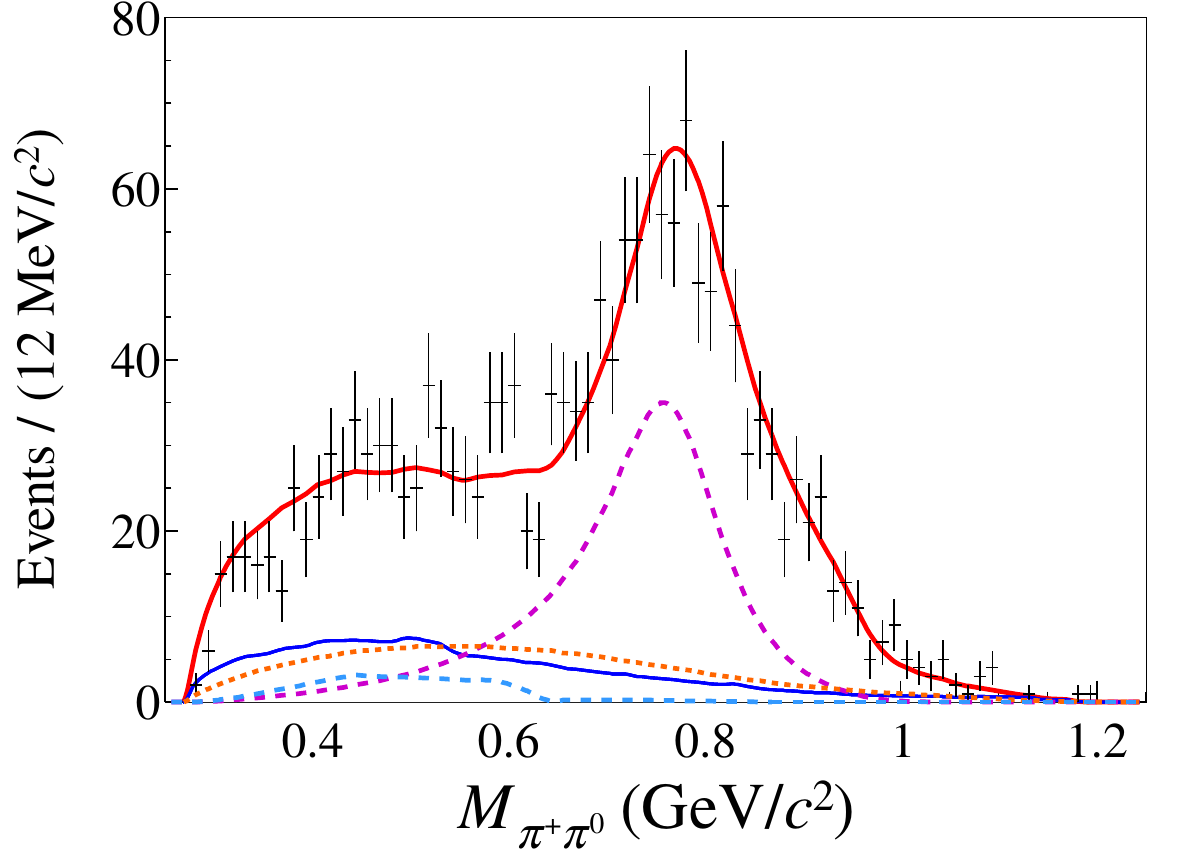}
  \includegraphics[width=4.2cm]{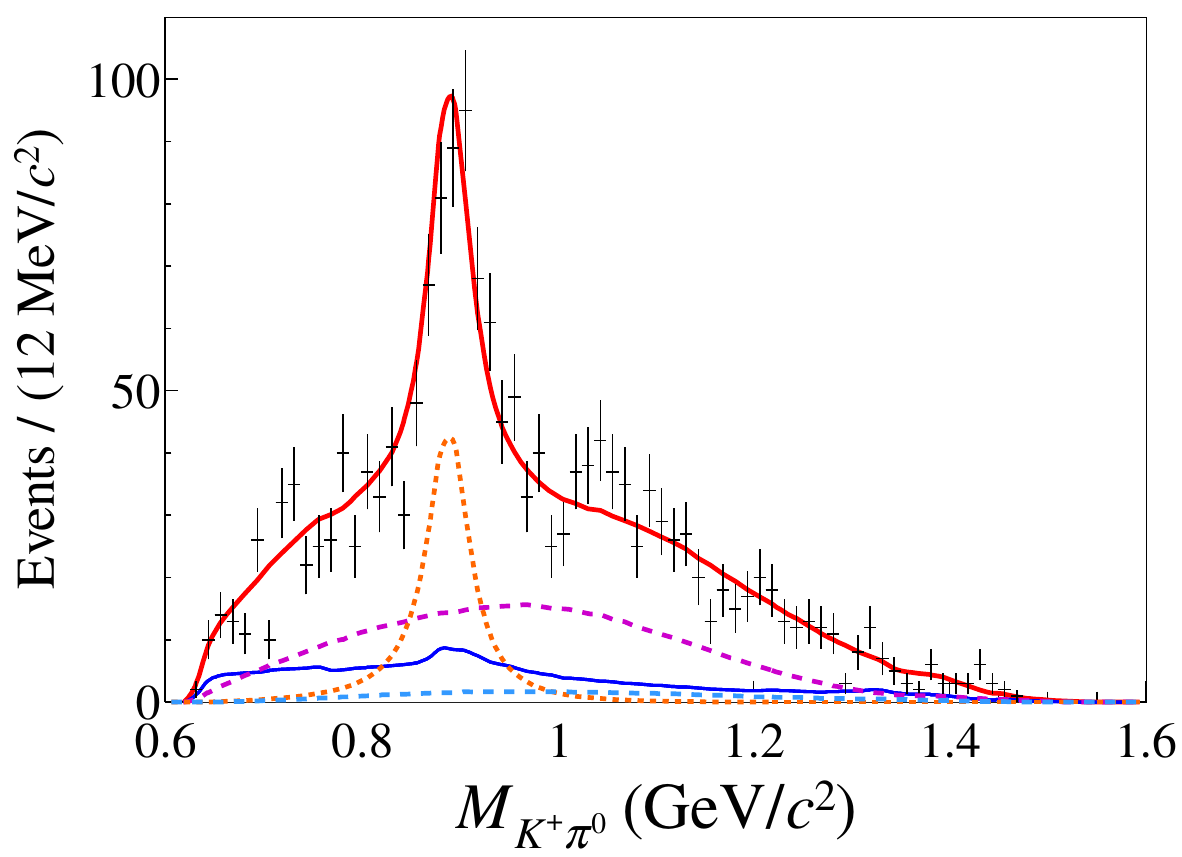}
  \includegraphics[width=4.2cm]{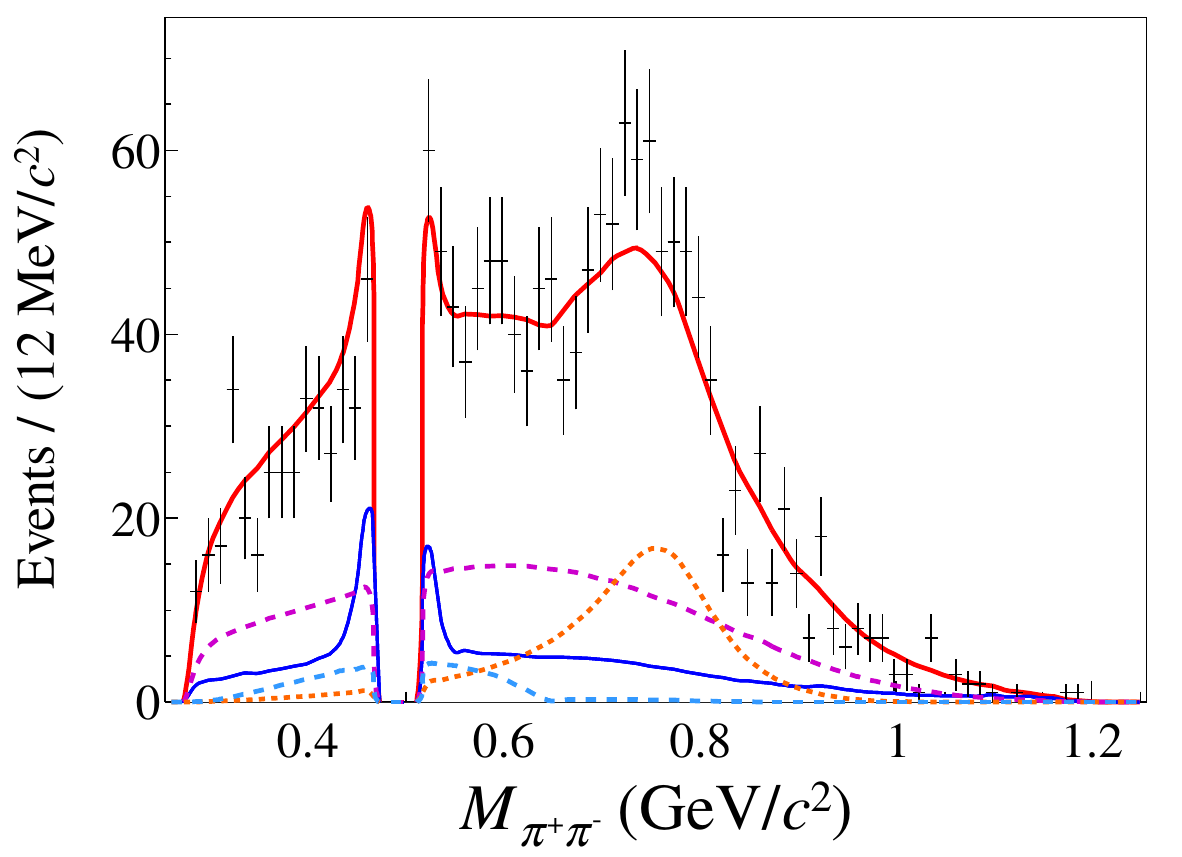}
  \includegraphics[width=4.2cm]{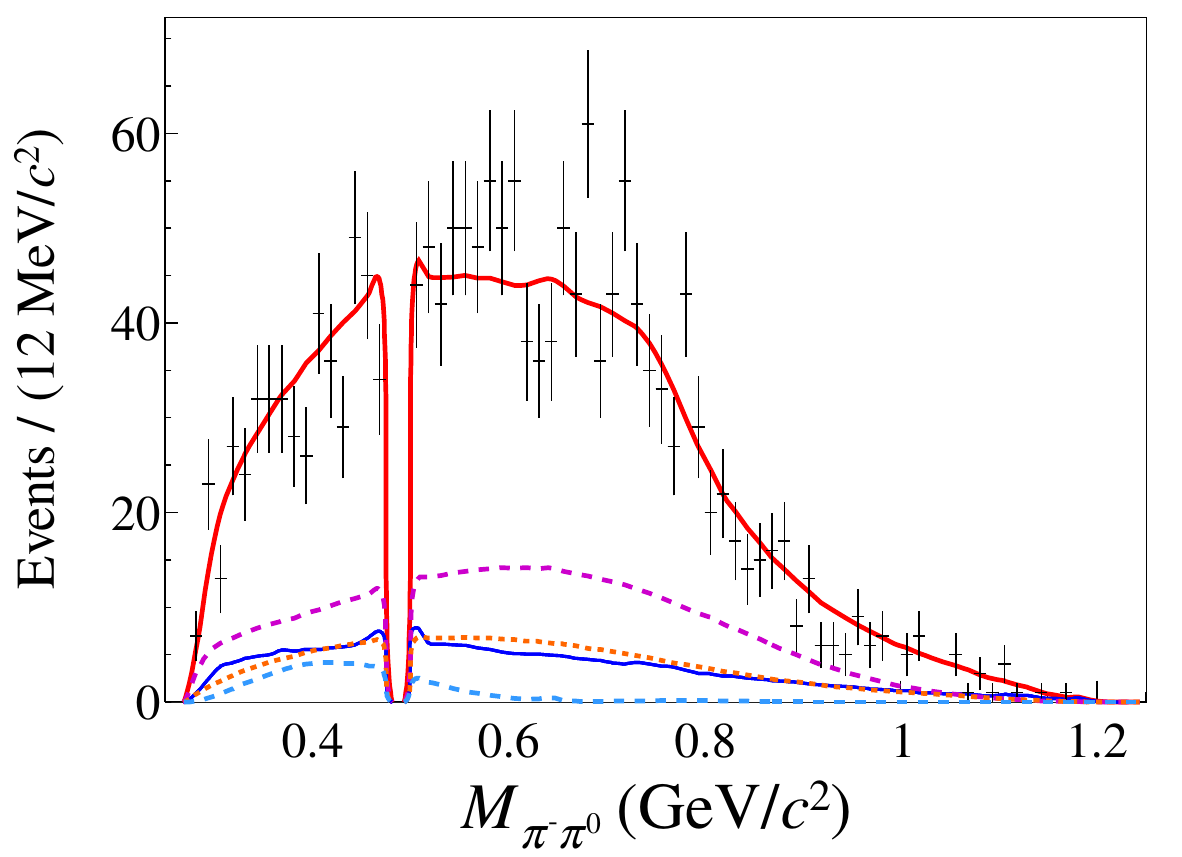}
  \includegraphics[width=4.2cm]{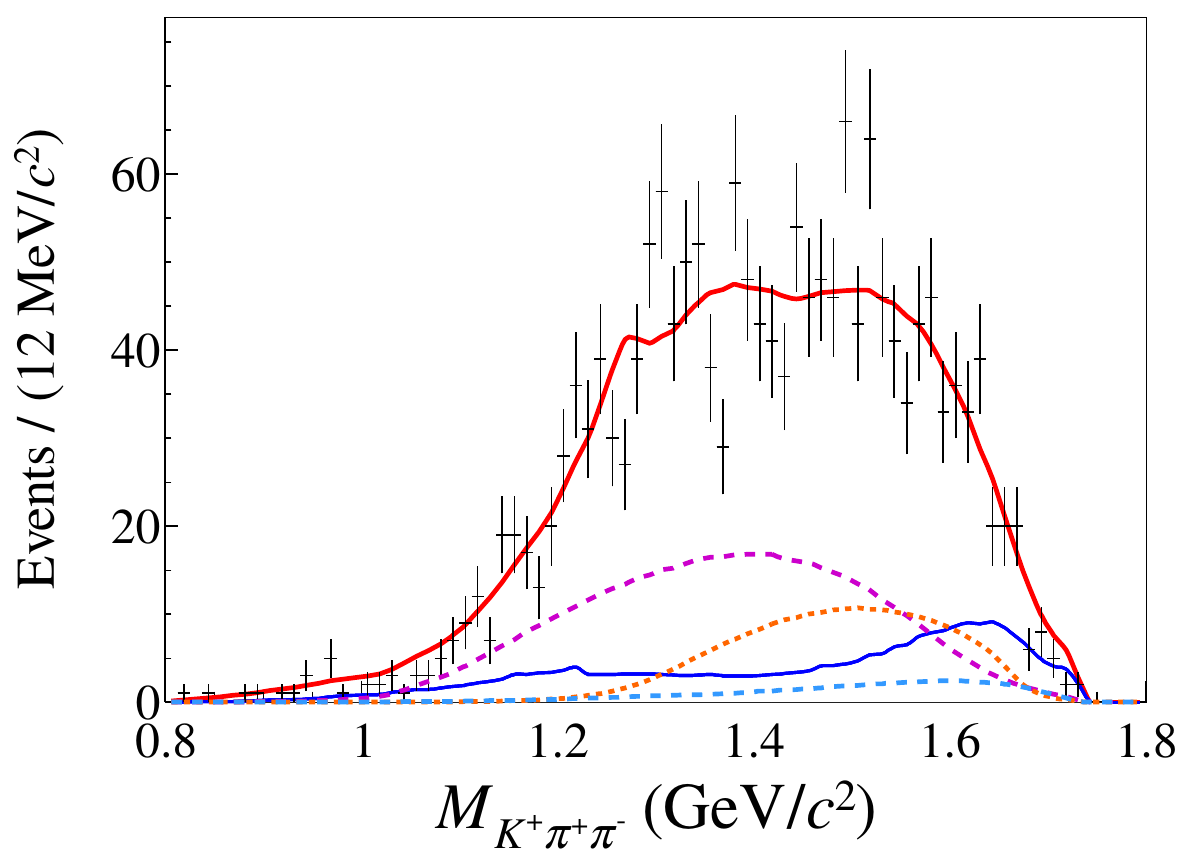}
  \includegraphics[width=4.2cm]{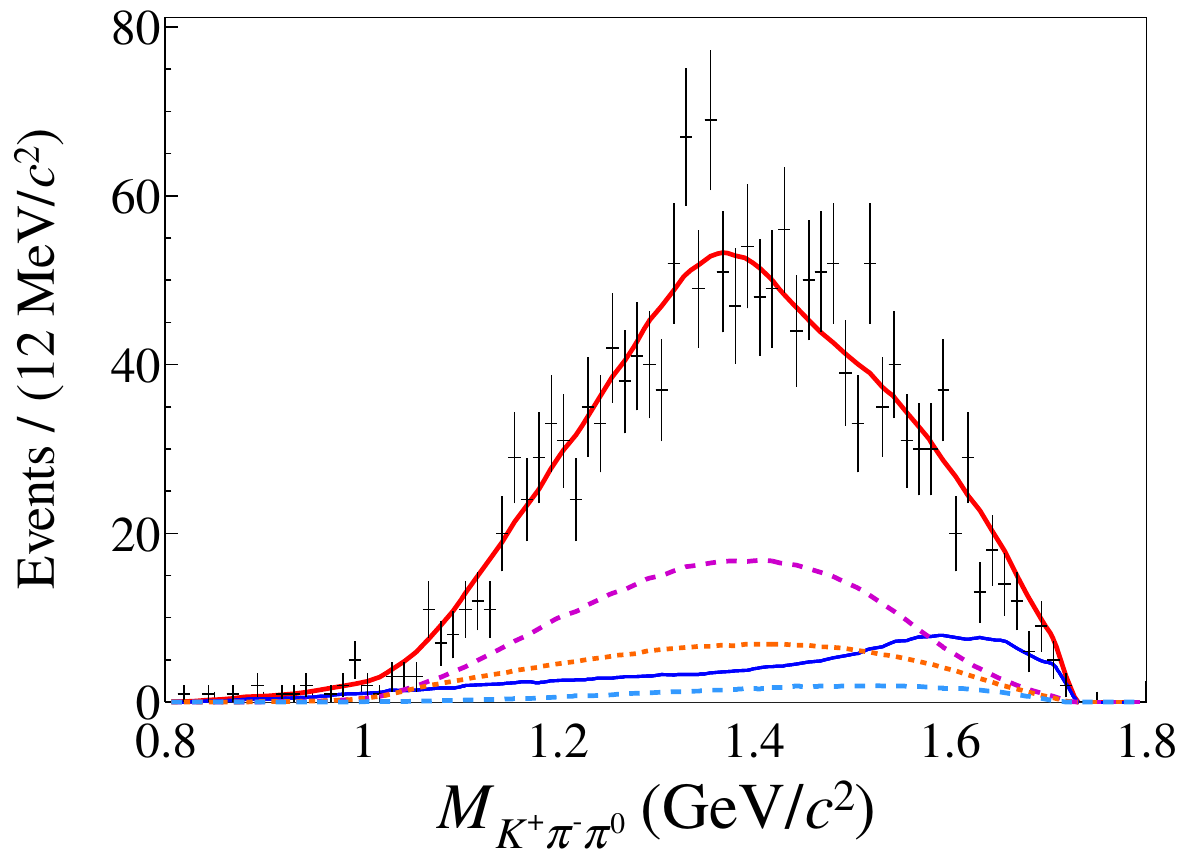}
  \includegraphics[width=4.2cm]{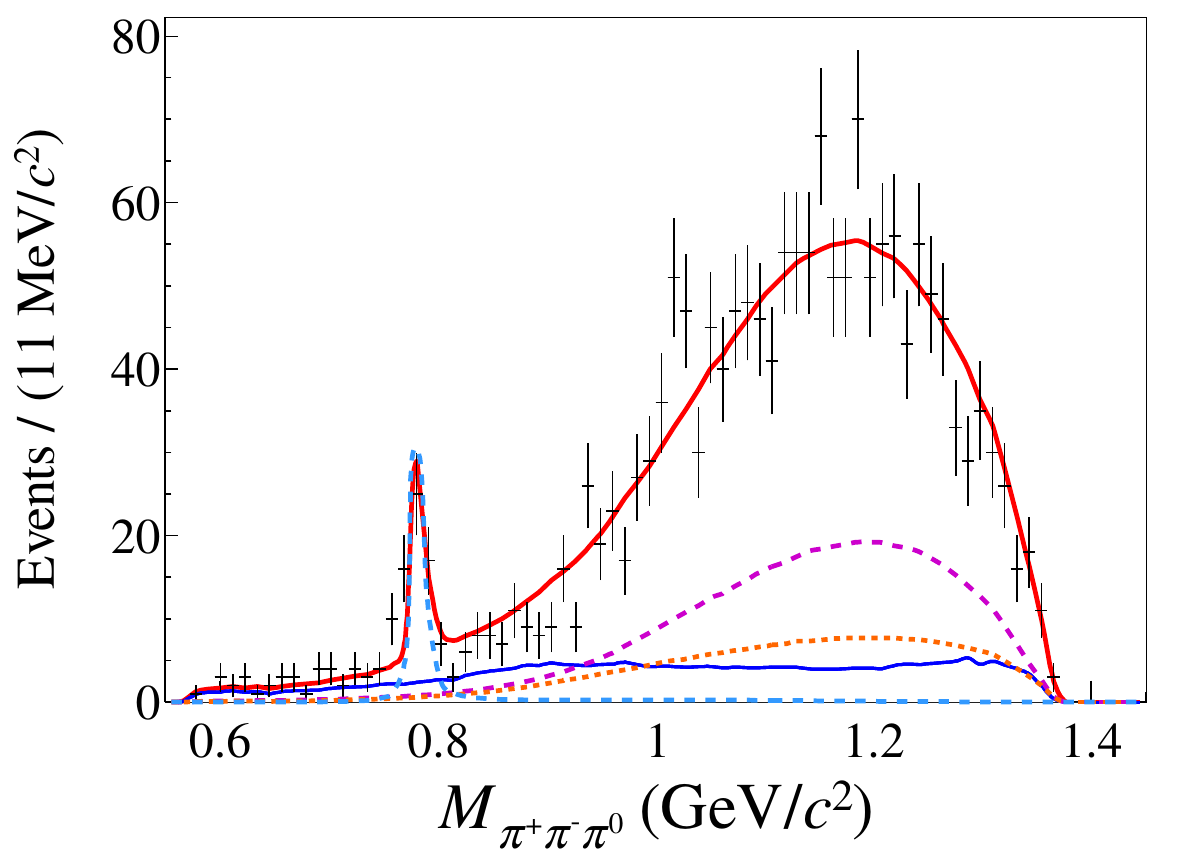}  
    \caption{Projections of the amplitude fit onto various invariant masses. The data are represented by points with error bars, the fit results by the red line, and the background by the blue line. The dashed lines with different colors represent intermediate processes.}
    \label{pwa:fitresult}
  \end{center}
\end{figure}

The initial amplitude model is constructed with the intermediate processes which are clearly evident in the invariant mass projections, including $K^{*0}\rho^+$, $K^{*+}\rho^0$ and $K^+\omega$. In the fit, the magnitude and phase of the process $D^+[\mathcal{S}]\to K^{*0}\rho^+$ are fixed to be one and zero, respectively, as reference. The coefficients of the sub-decays of $K_1(1400)$ and $a_1(1260)$ are related by Clebsch-Gordan coefficients according to isospin symmetry. 
Then all the possible combinations with different intermediate processes are tested and the model including the processes with statistical significance larger than $5\sigma$ is kept, where the statistical significance of each amplitude is calculated based on the change of the log-likelihood with and without that amplitude, taking into account the change of the degrees of freedom. 
The final model contains eleven components; the invariant mass projections are shown in Fig.~\ref{pwa:fitresult} and the phases, fit fractions (FFs) and BFs are listed in Table~\ref{tab:fitresult}. The FF of the $n^{th}$ amplitude is calculated by 
\begin{equation}
    {\rm FF}_n = \sum^{\rm N_{\rm MC}}{\left| \rho_n e^{i\phi_n}\mathcal{A}_n \right|^2} \,/\, \sum^{\rm N_{\rm MC}}{\left| \mathcal{M} \right|^2},
\end{equation}
where $\rm N_{\rm MC}$ is the number of phase space MC sample. The longitudinal polarization fraction, $F_L$ is obtained from the helicity amplitudes $H_{+}, H_0$, and $H_-$ which are related to the fitted covariant tensor coefficients.  The two measured fractions are  $F_L(D^+\to K^{*0}\rho^+)=0.111\pm0.024_{\rm stat}\pm0.008_{\rm syst}$ and $F_L(D^+\to K^{*+}\rho^0)=0.263\pm0.049_{\rm stat}\pm0.015_{\rm syst}$. Further details on $F_L$ and the interference between amplitudes can be found in the Supplemental Material~\cite{Supplement}. 
\nolinenumbers

\begin{table*}[tp!]
  \caption{Phases, FFs and BFs for each process of $D^+\to K^+\pi^+\pi^-\pi^0$. The first and the second uncertainties are statistical and systematic, respectively. The subscripts $A$ and $S$ represent the non-resonant axial-vector and scalar combinations of the two particles within parentheses. The BFs contain the sub-decay ratios. } 
  \centering
  \setlength{\tabcolsep}{16.5pt}
  \begin{tabular}{l| r@{$\pm$}c@{$\pm$}l |r@{$\pm$}c@{$\pm$}l |r@{$\pm$}c@{$\pm$}l}
    \hline
    \hline
    Amplitude  &\multicolumn{3}{c}{Phase $\phi_n$~(rad)} \vline &\multicolumn{3}{c}{FF (\%)} \vline &\multicolumn{3}{c}{BF ($\times10^{-4}$)} \vline\\
    \hline	
    $D^+[\mathcal{S}]\to K^{*0}\rho^+$
    &\multicolumn{3}{c}{0.0(fixed)} \vline   &14.9 &2.3 &0.8  &1.58  &0.25  &0.09 \\
    $D^+[\mathcal{P}]\to K^{*0}\rho^+$
    & 2.19 &0.09 &0.02  &17.1 &1.5 &0.4 &1.82  &0.17  &0.05 \\
    $D^+[\mathcal{D}]\to K^{*0}\rho^+$
    & 3.84 &0.09 &0.02  & 8.4 &1.5 &0.4 &0.89  &0.16  &0.04 \\
    $D^+\to K^{*0}\rho^+$ total
    &\multicolumn{3}{c}{-}  \vline  &35.6 &2.4 &1.0 &3.78  &0.27  &0.11 \\
    \hline
    $D^+[\mathcal{S}]\to K^{*+}\rho^0$
    & 2.62 &0.15 &0.04  & 4.5 &1.1 &0.3 &0.48  &0.12  &0.03 \\
    $D^+[\mathcal{P}]\to K^{*+}\rho^0$
    & 5.24 &0.12 &0.05  & 9.5 &1.1 &0.9 &1.01  &0.12  &0.10 \\
    $D^+[\mathcal{D}]\to K^{*+}\rho^0$
    & 1.08 &0.17 &0.02  & 2.6 &0.9 &0.2 &0.28  &0.10  &0.02 \\
    $D^+\to K^{*+}\rho^0$ total
    &\multicolumn{3}{c}{-}  \vline  &16.7 &1.7 &0.7 &1.77  &0.19  &0.08 \\
    \hline
    $D^+\to K^+a_1(1260)^0[\mathcal{S}](\to \rho^\pm \pi^\mp)$ 
    & 0.47 &0.16 &0.14  & 4.0 &1.1 &0.6 &0.43  &0.12  &0.06 \\
    \hline
    $D^+\to K^+a_1(1260)^0(\to f_0(500) \, \pi^0)$ 
    & 2.01 &0.16 &0.15  & 4.4 &1.2 &0.4 &0.47  &0.13  &0.04 \\
    \hline
    $D^+\to K_1(1270)^+[\mathcal{S}](\to K^{+}\rho^0) \, \pi^0$
    & 5.40 &0.17 &0.06  & 3.3 &0.8 &0.3 &0.35  &0.09  &0.03 \\
    \hline
    $D^+\to K_1(1400)^0[\mathcal{S}](\to K^{*}\pi) \, \pi^+$
    & 1.62 &0.17 &0.09  & 2.8 &0.9 &0.2 &0.30  &0.10  &0.02 \\
    \hline
    $D^+\to K_1(1400)^+[\mathcal{S}](\to K^{*0}\pi^+) \, \pi^0$
    & 4.72 &0.17 &0.09  & 3.5 &0.9 &0.2 &0.37  &0.10  &0.02 \\
    \hline
    $D^+\to K^+\omega$
    & 3.67 &0.38 &0.14  & 4.0 &0.7 &0.1 &0.43  &0.08  &0.01 \\
    \hline
    $D^+\to K^{*+} f_2(1270)$
    & 4.87 &0.18 &0.05  & 2.0 &0.8 &0.2 &0.21  &0.09  &0.02 \\
    \hline
    $D^+\to (K^+ \rho^-)_A \, \pi^0$
    & 0.18 &0.17 &0.08  & 6.6 &1.1 &0.4 &0.70  &0.12  &0.04 \\
    \hline
    $D^+\to (K^+\pi^-)_S \, (\pi^+\pi^0)_S$
    & 2.11 &0.19 &0.05  & 2.2 &0.7 &0.2 &0.23  &0.08  &0.02 \\
    \hline
    \hline
  \end{tabular}
  \label{tab:fitresult}
\end{table*}

The systematic uncertainties for the amplitude analysis from various sources are evaluated by assigning the difference between alternative fits and the nominal fit. The systematic uncertainty related to resonances, including the masses, widths and the line shape of $f_0(500)$ is estimated by varying their parameters within uncertainties~\cite{pdg} and changing with the RBW function, respectively. The barrier radii for $D^+$ and the intermediate states are varied by $\pm1\ ({\rm GeV}/c)^{-1}$. The uncertainty associated with the detector acceptance difference between MC samples and data is determined by re-weighting the MC events in the likelihood function according to the acceptance difference. The uncertainty related to the purity is studied by varying the purity within its statistical uncertainty, and the background shape is studied by varying the proportion of the continuum light hadron background by $\pm5\%$~\cite{BES:2007cev}. The fitter performance has been checked by fitting data-sized simulation samples generated with the amplitude models that result from the studies. All the fitted pull values are assigned as the corresponding systematic uncertainties. The uncertainty from insignificant processes (i.e., low-significance amplitudes) is considered by including the process with the largest significance. The total uncertainties are determined by adding all the contributions in quadrature. Detailed results can be found in the Supplemental Material~\cite{Supplement}.

The same tag modes and selection criteria as the amplitude are employed to perform the branching fraction measurement. The DT yields ($Y_{\rm DT}$) is determined to be $1654\pm46$ in the whole $M_{\rm BC}$ region. With the ST yields ($Y_{\rm ST}$) and efficiencies ($\epsilon_{\rm ST}$) cited by Ref.~\cite{BESIII:2025vbt}, and the DT efficiencies ($\epsilon_{\rm DT}$) estimated with the corresponding signal MC samples, the BF can be determined by $\mathcal{B}(D^+ \to K^+\pi^+\pi^-\pi^0) = Y_{\rm DT}/\sum_i Y_{\rm ST}^i (\epsilon_{\rm DT}^i/\epsilon_{\rm ST}^i)$ , where the index $i$ denotes the $i^{th}$ tag mode. The fit result is shown in Fig.~\ref{BF:fitresult}. The BF is measured to be $(1.062\pm0.028_{\rm stat}\pm0.012_{\rm syst})\times10^{-3}$.

The systematic uncertainty of the ST yield is estimated as 0.3\%, based on varying the signal shape, the background shape and leaving the parameters of one smearing Gaussian free in the $M_{\rm BC}^{\rm tag}$ fit. The tracking and particle identification efficiencies of $\pi^\pm$, $K^\pm$ as well as the $\pi^0$ reconstruction efficiency are studied with DT $D\bar{D}$ hadronic events. After correcting the MC efficiencies to those found in data, the uncertainties of the correction parameters are assigned as the corresponding uncertainties.  These are
$0.3\%, 0.3\%$, and $0.2\%$ for tracking, PID, and $\pi^0$ reconstruction, respectively.  
The uncertainty associated with the amplitude model is estimated by varying the fitted parameters based on the covariance matrix and randomizing the fixed parameters by a Gaussian distribution. The distribution of 600 efficiencies arising from this procedure is fitted by a Gaussian function and the deviation from the nominal mean value, 0.5\%, is taken as the systematic uncertainty. To estimate the systematic uncertainty from the 2D fit, the measurements are repeated by varying the signal shape, the endpoint of the ARGUS function, and the fixed number of background events from $D^+\to K^+K_S^0\pi^0$ and $D^+\to K^+K^-\pi^+$. Quadratically summing over the changes of the BF gives the systematic uncertainty, which is 0.6\%. The systematic uncertainty associated with the $\Delta E_{\rm sig}$ requirement is evaluated to be 0.6\%, estimated by smearing the $\Delta E_{\rm sig}$ distribution for signal MC events. The total systematic uncertainty of the BF measurement is 1.1\%.

In summary, the first amplitude analysis of the DCS decay $D^+\to K^+\pi^+\pi^-\pi^0$ is performed using 20.3~fb$^{-1}$ of data taken at $\sqrt{s}=3.773$~GeV with the BESIII detector.
The BFs with and without the contribution from $D^+\to K^+\omega$ are measured to be $(1.062\pm0.028_{\rm stat}\pm0.012_{\rm syst})\times10^{-3}$ and $(1.010\pm0.029_{\rm stat}\pm0.013_{\rm syst})\times10^{-3}$, respectively. The latter is consistent with the previous BESIII results within 1.2$\sigma$~\cite{BESIII:2020wnc,BESIII:2021uix}.

The $VV$ final states $D^+\to K^{*0}\rho^+$ and $D^+\to K^{*+}\rho^0$ are found to dominate this decay, accounting for about half of the total FFs. The corresponding BFs are $(5.67\pm0.41_{\rm stat}\pm0.17_{\rm syst})\times10^{-4}$ and $(5.32\pm0.57_{\rm stat}\pm0.24_{\rm syst})\times10^{-4}$, respectively. It is evident that the internal structure differs from that of the corresponding CF decay, $D^+\to K^-\pi^+\pi^+\pi^0$, which is dominated by $D^+\to\bar{K}^{*0}\rho^+$, leading to a deviation of the DCS/CF ratio from tan$^4\theta_C$~\cite{BESIII:2024waz}.
The BF of $D^+\to K^+\omega$ is measured to be $(4.76\pm0.84_{\rm stat}\pm0.13_{\rm syst})\times 10^{-5}$, consistent with previous measurement~\cite{BESIII:2020wnc} but significantly smaller than all predictions based on the topological diagram approach~\cite{Cheng:2016ejf,Cheng:2024hdo}, the factorization-assisted topological-amplitude approach~\cite{Qin:2013tje}, and the pole model~\cite{Fu-Sheng:2011fji}.
These BFs provide important inputs for testing whether the dynamical mechanisms of CF and DCS decays are consistent, and for investigating SU(3) symmetry breaking effects.

Furthermore, the two $D\to VV$ processes are both found to be mostly transversely polarized. Their longitudinal fractions are measured to be $F_L(D^+\to K^{*0}\rho^+)=0.111\pm0.024_{\rm stat}\pm0.008_{\rm syst}$ and $F_L(D^+\to K^{*+}\rho^0)=0.263\pm0.049_{\rm stat}\pm0.015_{\rm syst}$, respectively. 
The corresponding $\mathcal{S}$-wave components are much smaller than that in the CF decay $D^+\to \bar{K}^{*0}\rho^+$~\cite{BESIII:2023qgj,BESIII:2024waz}, which also contradicts the naive expectation of $\mathcal{S}$-wave dominance. However, in $D^0$ decays, the CF $D^0\to\bar{K}^{*0}\rho^0$ and the DCS $D^0\to{K}^{*0}\rho^0$ decays exhibit identical polarization behavior~\cite{LHCb:2017swu}. Additionally, a similar pattern of $\mathcal{P}$-wave dominance appears in the decays $D_s^+\to K^{*}\rho$~\cite{BESIII:2022bvv}, which are singly Cabibbo-suppressed and share analogous weak topologies with the two DCS $D^+\to K^*\rho$ decays. The polarization behavior in DCS sector provides critical insights into resolving the long-standing polarization puzzle, particularly for probing the contribution from long-distance final-state interactions~\cite{Cao:2023csx,Cheng:2024hdo,Yu:2024vlz}.

The authors thank Professor Hai-Yang Cheng and Professor Yu-Kuo Hsiao for helpful discussions. The BESIII Collaboration thanks the staff of BEPCII (https://cstr.cn/31109.02.BEPC) and the IHEP computing center for their strong support. This work is supported in part by National Key R\&D Program of China under Contracts Nos. 2023YFA1606000, 2023YFA1606704; National Natural Science Foundation of China (NSFC) under Contracts Nos. 12635006, 12675104, 12035009, 11635010, 11935015, 11935016, 11935018, 12025502, 12035013, 12061131003, 12192260, 12192261, 12192262, 12192263, 12192264, 12192265, 12221005, 12225509, 12235017, 12342502, 12361141819; Joint Large-Scale Scientific Facility Fund of the NSFC and the Chinese Academy of Sciences under Contract No.~U2032104; the Chinese Academy of Sciences (CAS) Large-Scale Scientific Facility Program; the Strategic Priority Research Program of Chinese Academy of Sciences under Contract No. XDA0480600; the Excellent Youth Foundation of Henan Scientific Committee under Contract No.~242300421044; CAS under Contract No. YSBR-101; 100 Talents Program of CAS; The Institute of Nuclear and Particle Physics (INPAC) and Shanghai Key Laboratory for Particle Physics and Cosmology; ERC under Contract No. 758462; German Research Foundation DFG under Contract No. FOR5327; Istituto Nazionale di Fisica Nucleare, Italy; Knut and Alice Wallenberg Foundation under Contracts Nos. 2021.0174, 2021.0299, 2023.0315; Ministry of Development of Turkey under Contract No. DPT2006K-120470; National Research Foundation of Korea under Contract No. NRF-2022R1A2C1092335; National Science and Technology fund of Mongolia; Polish National Science Centre under Contract No. 2024/53/B/ST2/00975; STFC (United Kingdom); Swedish Research Council under Contract No. 2019.04595; U. S. Department of Energy under Contract No. DE-FG02-05ER41374.

\bibliography{references}

\onecolumngrid

\section{{\bf \boldmath Supplemental Material: Observation of $D^+ \to K^{*0}\rho^+$ and $D^+\to K^{*+}\rho^0$ in Doubly Cabibbo-Suppressed Decay $D^+ \to K^+\pi^+\pi^-\pi^0$}}

\subsection{1. Longitudinal polarization fraction for $D\to VV$}

To calculate the longitudinal polarization fraction $F_L$, the helicity amplitude for $D^+\to V_1V_2\to(P_1P_2)(P_3P_4)$ processes is constructed as:
\begin{equation}
    H_{\lambda_{V_1},\lambda_{V_2}}=F_{\lambda_{V_1},\lambda_{V_2}}^{J_D}D_{\lambda_{D},\lambda_{V_1}-\lambda_{V_2}}^{J_D}(\phi_{V_1},\theta_{V_1},0) \,  P_{V_1}F_{\lambda_{P_1},\lambda_{P_2}}^{J_{V_1}}D_{\lambda_{V_1},\lambda_{P_1}-\lambda_{P_2}}^{J_{V_1}}(\phi_{P_1},\theta_{P_1},0) \,  P_{V_2}F_{\lambda_{P_3},\lambda_{P_4}}^{J_{V_2}}D_{\lambda_{V_2},\lambda_{P_3}-\lambda_{P_4}}^{J_{V_2}}(\phi_{P_3},\theta_{P_3},0),\\
\end{equation}
where $\lambda$ denotes the helicity of particle, which is rotationally invariant. The superscript $J$ indicates the particle spin, and $\theta$ and $\phi$ are the helicity angles. The $P_{V_{1,2}}$ factor is the propagator for vector mesons, $D$ is the Wigner D-function, and $F_{\lambda_1,\lambda_2}$ is the helicity-coupling amplitude: 
\begin{equation}
    F_{\lambda_1,\lambda_2}^{A\to BC} = \sum_{LS} \, \alpha_{LS} \, \sqrt{\frac{2L+1}{2J_A+1}} \,\langle L0;S\delta|J_A\delta\rangle \, \langle J_BJ_C;\lambda_1 -\lambda_2|S\delta\rangle \, q^LX_L.  \\
    \label{eq:helicity}
\end{equation}
Here, $L$ denotes the relative angular momentum between daughter particles $B$ and $C$, $S$ is the coupled spin. The helicity difference is $\delta=\lambda_1-\lambda_2$, $q$ is the decay momentum and $X_L$ denotes the Blatt-Weisskopf barrier. 
For this specific process, $J_D=0$, and $\lambda_{V_1}=\lambda_{V_2}$, which lead to three helicity states: $H_+$, $H_0$ and $H_-$. Therefore, the total amplitude is $|A|^2=|H_++H_0+H_-|^2$. The helicity amplitude coefficients for $D\to VV$ also have three components $\alpha_{0,1,2}$ corresponding to the angular momentum between vector mesons. Based on the above description, we obtain:
\begin{equation}
\begin{gathered}
    H_{\pm} = \left( \sqrt{\frac{1}{3}}\alpha_0\mp\sqrt{\frac{1}{2}}\alpha_1M_{D}q_DX_1^D+\sqrt{\frac{1}{6}}\alpha_2q_D^2X_2^D \right) \, e^{\pm\phi_{P_1}}\frac{\sin\theta_{P_1}}{\sqrt{2}}e^{\pm\phi_{P_3}}\frac{\sin\theta_{P_3}}{\sqrt{2}}q_{V_1}X_1^{V_1}q_{V_2}X_1^{V_2}P_{V_1}P_{V_2},\\
    H_{0} = \left( -\sqrt{\frac{1}{3}}\alpha_0+\sqrt{\frac{2}{3}}\alpha_2q_D^2X_2^D \right) \, \gamma_{V_1}\gamma_{V_2}\cos\theta_{P_1}\cos\theta_{P_3}q_{V_1}X_1^{V_1}q_{V_2}X_1^{V_2}P_{V_1}P_{V_2}.
\end{gathered}
\end{equation}
Here, $M_D$ is the mass of $D^+$ meson, $\gamma$ is the Lorentz factor. With the following formulas, the coefficients of covariant tensor amplitude $c_{0,1,2}$ can be converted to the helicity amplitude coefficients by:
\begin{equation}
    \alpha_0=4\sqrt{3}c_0, \quad
    \alpha_1=8\sqrt{2} \, e^{-i\pi/2}c_1, \quad
    \alpha_2=16\sqrt{\frac{2}{3}}e^{-i\pi}c_2.\\
\end{equation}
With the coefficients $c_{0,1,2}$ obtained from the amplitude analysis, the helicity amplitudes $H_{\pm}$ and $H_0$ are obtained for each event. The longitudinal polarization fraction $F_L$ is defined as:
\begin{equation}
    F_L=\frac{\int|H_0|^2d\Phi}{\int|H_-+H_0+H_+|^2d\Phi}=\frac{\sum_{\rm MC}|H_0|^2}{\sum_{\rm MC}|H_-+H_0+H_+|^2}.
    \label{eq:fl}
\end{equation}
The integral in Eq.~(\ref{eq:fl}) is achieved by using phase space Monte-Carlo (MC) events without detection efficiency. The statistical uncertainty for $F_L$ is estimated by varying the coefficients $c_{0,1,2}$ according to the correlation matrix. A Gaussian function is then employed to fit the $F_L$ distribution and the width is assigned as the uncertainty.

\subsection{2. Systematic uncertainties in amplitude analysis}

The detailed values of the systematic uncertainties of amplitude analysis are listed. The following six sources are considered: amplitude model, data-MC differences, background estimation, barrier radii, fit bias and insignificant amplitudes. The total uncertainties are determined by adding all the contributions in quadrature. All the systematic uncertainties are listed in Table~\ref{tab:pwa_sys}.

\begin{table}[htbp]
  \caption{Systematic uncertainties on the phases and FFs for different amplitudes in units of the corresponding statistical uncertainties. Groups of related amplitudes are separated by horizontal lines. The last row of the first two groups gives the corresponding uncertainties of the total fit fraction of the above components, including interference effects.\\
    (i) amplitude model, (ii) data-MC differences, (iii) background estimation, (iv) barrier radii, (v) fit bias, (vi) insignificant amplitudes. }
  \centering
\vspace{1em}
  \begin{tabular}{clcccccccc}
    \hline
    \hline
      &\multicolumn{8}{c}{Source}    & \\
    \hline
    &Amplitude  &\  &i  &ii  &iii  &iv &v &vi  &Total\\
    \hline
    I &$D^+[S]\to K^{*0}\rho^+$
    &FF     &0.21 &0.06 &0.11 &0.06 &0.06 &0.21 &0.33  \\ 
    \multirow{2}{*}{II} &\multirow{2}{*}{$D^+[P]\to K^{*0}\rho^+$}
    &$\phi$ &0.07 &0.01 &0.01 &0.21 &0.01 &0.05 &0.23  \\
    & &FF     &0.10 &0.05 &0.12 &0.20 &0.02 &0.04 &0.26  \\ 
    \multirow{2}{*}{III} &\multirow{2}{*}{$D^+[D]\to K^{*0}\rho^+$}
    &$\phi$ &0.12 &0.02 &0.03 &0.16 &0.04 &0.12 &0.24  \\
    & &FF     &0.09 &0.02 &0.09 &0.02 &0.08 &0.25 &0.29  \\ 
    &$D^+\to K^{*0}\rho^+$ total
    &FF     &0.23 &0.08 &0.21 &0.08 &0.06 &0.26 &0.43  \\
    \hline 
    \multirow{2}{*}{IV} &\multirow{2}{*}{$D^+[S]\to K^{*+}\rho^0$}
    &$\phi$ &0.16 &0.04 &0.10 &0.17 &0.02 &0.10 &0.28  \\
    & &FF     &0.14 &0.03 &0.06 &0.07 &0.00 &0.17 &0.24  \\ 
    \multirow{2}{*}{V} &\multirow{2}{*}{$D^+[P]\to K^{*+}\rho^0$}
    &$\phi$ &0.08 &0.00 &0.01 &0.30 &0.01 &0.22 &0.38  \\
    & &FF     &0.07 &0.05 &0.02 &0.19 &0.04 &0.76 &0.79  \\ 
    \multirow{2}{*}{VI} &\multirow{2}{*}{$D^+[D]\to K^{*+}\rho^0$}
    &$\phi$ &0.08 &0.03 &0.02 &0.08 &0.02 &0.07 &0.14  \\
    & &FF     &0.05 &0.02 &0.01 &0.16 &0.04 &0.02 &0.17  \\ 
    &$D^+\to K^{*+}\rho^0$ total
    &FF     &0.05 &0.03 &0.01 &0.05 &0.03 &0.39 &0.40  \\ 
    \hline 
    \multirow{2}{*}{VII} &\multirow{2}{*}{$D^+\to K^+a_1(1260)^0[S](\to \rho^{\pm}\pi^{\mp})$}
    &$\phi$ &0.84 &0.01 &0.04 &0.04 &0.02 &0.08 &0.85  \\
    & &FF     &0.23 &0.05 &0.13 &0.01 &0.07 &0.47 &0.55  \\ 
    \hline 
    \multirow{2}{*}{VIII} &\multirow{2}{*}{$D^+\to K^+a_1(1260)^0[S](\to f_0(500)\pi^{0})$}
    &$\phi$ &0.83 &0.01 &0.04 &0.07 &0.06 &0.37 &0.91  \\
    & &FF     &0.21 &0.06 &0.02 &0.18 &0.00 &0.13 &0.31  \\ 
    \hline 
    \multirow{2}{*}{IX} &\multirow{2}{*}{$D^+\to K_1(1270)^+[S](\to K^{+}\rho^0)\pi^0$}
    &$\phi$ &0.13 &0.04 &0.08 &0.24 &0.04 &0.20 &0.35  \\
    & &FF     &0.09 &0.01 &0.02 &0.20 &0.01 &0.23 &0.32  \\ 
    \hline 
    \multirow{2}{*}{X} &\multirow{2}{*}{$D^+\to K_1(1400)^0[S](\to K^{*}\pi)\pi^+$}
    &$\phi$ &0.37 &0.02 &0.05 &0.02 &0.04 &0.33 &0.50  \\
    & &FF     &0.17 &0.04 &0.02 &0.16 &0.04 &0.06 &0.25  \\ 
    \hline 
    \multirow{2}{*}{XI} &\multirow{2}{*}{$D^+\to K_1(1400)^+[S](\to K^{*0}\pi^+)\pi^0$}
    &$\phi$ &0.40 &0.01 &0.09 &0.02 &0.01 &0.36 &0.55  \\
    & &FF     &0.17 &0.01 &0.03 &0.16 &0.06 &0.05 &0.25  \\ 
    \hline
    \multirow{2}{*}{XII} &\multirow{2}{*}{$D^+\to K^+\omega$}
    &$\phi$ &0.09 &0.04 &0.05 &0.24 &0.08 &0.25 &0.37  \\
    & &FF     &0.03 &0.06 &0.04 &0.08 &0.03 &0.10 &0.15  \\ 
    \hline
    \multirow{2}{*}{XIII} &\multirow{2}{*}{$D^+\to K^{*+} f_2(1270)$}
    &$\phi$ &0.07 &0.01 &0.02 &0.21 &0.05 &0.17 &0.28  \\
    & &FF     &0.07 &0.01 &0.06 &0.16 &0.07 &0.11 &0.23  \\ 
    \hline
    \multirow{2}{*}{XIV} &\multirow{2}{*}{$D^+\to (K^+ \rho^-)_A \pi^0$}
    &$\phi$ &0.27 &0.03 &0.12 &0.13 &0.06 &0.34 &0.47  \\
    & &FF     &0.10 &0.01 &0.06 &0.26 &0.04 &0.18 &0.34  \\ 
    \hline
    \multirow{2}{*}{XV} &\multirow{2}{*}{$D^+\to (K^+\pi^-)_S(\pi^+\pi^0)_S$}
    &$\phi$ &0.07 &0.04 &0.09 &0.19 &0.02 &0.07 &0.24  \\
    & &FF     &0.04 &0.07 &0.09 &0.30 &0.02 &0.07 &0.33  \\ 
    \hline    \hline
    &$F_L(D^+\to K^{*0}\rho^+)$
    &       &0.09 &0.02 &0.04 &0.24 &0.03 &0.20 &0.33 \rule{0pt}{11pt} \\
    \hline
    &$F_L(D^+\to K^{*+}\rho^0)$
    &       &0.07 &0.08 &0.03 &0.20 &0.04 &0.21 &0.31 \rule{0pt}{11pt} \\
    \hline
    \hline
  \end{tabular}
  \label{tab:pwa_sys}
\end{table}



\subsection{3. Interference between the amplitudes}


The interference fit fraction IN between the $n^{\rm th}$ and $n^{\prime {\rm th}}$ amplitudes is defined as
\begin{eqnarray}\begin{aligned}
    {\rm IN}_{nn^{\prime}} = \frac{\sum^{N_{\rm gen}} 2{\rm Re}[c_{n}c^{*}_{n^{\prime}}\mathcal{A}_{n}\mathcal{A}^{*}_{n^{\prime}}]}{\sum^{N_{\rm gen}} \left|\mathcal{M}\right|^{2}}, \label{interferenceFF-Definition}
\end{aligned}\end{eqnarray}
where $N_{\rm gen}$ is the number of phase space signal MC events at generator level, $c_n=\rho_n e^{i\phi_n}$.

 The interference fit fractions between the amplitudes are listed in Table~\ref{tab:inter}.

\begin{table}[htbp]
  \caption{Interference fit fractions between the amplitudes, in unit of \% of total amplitude. The Roman numeral amplitude labels are the same as Table~\ref{tab:pwa_sys}.}
  \centering
  \vspace{1em}
  \begin{tabular}{c|r r r r r r r r r r r r r r}
    \hline
    \hline
    &II &III &IV &V &VI &VII &VIII &IX &X &XI &XII &XIII &XIV &XV\\
    \hline
    I    & 0.0 &$-$4.9 &$-$1.2 & 0.0 & 0.0 & 2.4 & 3.1 & 0.6 & 1.0 & 1.9 & 0.0 &$-$0.3 & 0.7 & 0.0 \\
    II   & & 0.0 & 0.0 & 2.5 & 0.0 & 0.0 & 0.0 & 0.0 & 0.0 & 0.0 & 1.3 & 0.0 & 0.0 & 0.0\\
    III  & & & 0.0 & 0.0 &$-$0.5 & 0.1 &$-$1.7 &$-$0.1 &$-$1.1 &$-$1.7 & 0.0 &$-$0.2 & 0.3 & 0.0\\
    IV   & & & & 0.0 & 0.1 &$-$1.1 & 0.0 & 1.1 &$-$0.2 & 0.0 & 0.0 & 0.0 &$-$0.2 & 0.1\\
    V    & & & & & 0.0 & 0.0 & 0.0 & 0.0 & 0.0 & 0.0 & 1.0 & 0.0 & 0.0 & 0.0\\
    VI   & & & & & & 0.1 & 0.0 & 0.2 &$-$0.7 &$-$0.1 & 0.0 & 0.0 & 0.1 &$-$0.1\\
    VII  & & & & & & & 0.0 & 0.0 &$-$0.1 & 0.0 & 0.0 & 0.0 & 0.9 &$-$0.2\\
    VIII & & & & & & & & 0.0 & 0.2 & 0.5 & 0.0 & 0.0 & 1.5 & 0.2\\
    IX   & & & & & & & & & 0.0 & 0.7 & 0.0 & 0.0 &$-$0.9 & 0.0\\
    X    & & & & & & & & & & 0.4 & 0.0 & 0.7 & 1.4 & 0.0\\
    XI   & & & & & & & & & & & 0.0 & 0.0 & 0.1 & 0.0\\
    XII  & & & & & & & & & & & & 0.0 & 0.0 & 0.0\\
    XIII & & & & & & & & & & & & & 0.6 & 0.0\\
    XIV  & & & & & & & & & & & & & & 0.0\\
    \hline
    \hline
  \end{tabular}
  \label{tab:inter}
\end{table}


\subsection{4. Systematic uncertainties in the BF measurement}


The detailed values of systematic uncertainties in the BF measurement are listed. The following seven sources are considered: ST yield, tracking, particle identification (PID), $\pi^0$ reconstruction, amplitude model, 2D fit, and $\Delta E_{\rm sig}$ requirements. All the systematic uncertainties are listed in Table~\ref{BF-Sys}.

\begin{table}[htbp]
  \centering
      \caption{Systematic uncertainties in the BF measurement.}
  \begin{tabular}{c c}
  \hline
  \hline
    Source            &Sys. Uncertainty (\%) \\
    \hline
    ST yield                &0.3\\
    Tracking                &0.3\\
    PID                     &0.3\\
    $\pi^0$ reconstruction  &0.2\\
    Amplitude model         &0.5\\
    2D fit                  &0.6\\
    $\Delta E_{\rm sig}$ requirement  &0.6\\
    \hline
    Total                   &1.1\\
    \hline
    \hline
  \end{tabular}
  \label{BF-Sys}
\end{table}

\end{document}